\documentclass[a4paper, 10pt]{article}

\usepackage{graphicx}
\usepackage{amsfonts, amstext, amsmath}
\usepackage{subfigure}
\usepackage[a4paper]{geometry}

\newcommand{\R}{\mathbb{R}}
\newcommand{\eps}{\varepsilon}

\def \And{\;\text{ and }\;}

\title{Inflation and speculation in a dynamic macroeconomic model}

\author{ Matheus R. Grasselli\footnote{{\bf Aknowledgement:} the author thanks the participants of the Fifth Mathematic in Finance Conference (Kruger Park, South Africa, August 2014) where part of this work was presented. This research received partial financial support from the Institute for New Economic Thinking (Grant INO13-00011) and the Natural Sciences and Engineering Research Council of Canada (Discovery Grants)}
               \\ \small McMaster University
               \\ \small 1280 Main St. W, Hamilton, ON L8S 4L8, Canada
               \\ \small grasselli@math.mcmaster.ca 
               \\ \\ 
 Adrien Nguyen Huu
 			\\ \small \'Ecole Nationale des Ponts et Chaussées 
            \\ \small 6-8 Av. Blaise Pascal, 77455 Marne-la-Vall\'ee Cedex 2, France
            \\ \small adrien.nguyen-huu@enpc.fr}

\begin{document}

\maketitle

\begin{abstract}
We study a monetary version of the Keen model by merging
two alternative extensions, namely the addition of a dynamic price level
and the introduction of speculation. We recall and study old and new equilibria, together with their local stability analysis. This
includes a state of recession associated with a deflationary regime and characterized by falling employment but constant wage shares, with 
or without an accompanying debt crisis. We also emphasize some new qualitative behavior of the extended model, 
in particular its ability to produce and describe repeated financial crises as a natural pace of the economy,
and its suitability to describe the relationship between economic growth and financial activities.
\end{abstract}

\noindent {\bf Key words~:} 
Minsky's financial instability hypothesis;
Keen model;
stock-flow consistency;
financial crisis;
dynamical systems in macroeconomics;
local stability;
limit cycles

\section{Introduction}
\label{sec:intro}

More than six years after the end of the 2007-08 financial crisis, most advanced economies find themselves 
in regimes of low inflation or risk of deflation \cite{Hilsenrath_Blackstone2014}. Because the crisis itself is generally 
regarded as having speculative excesses as one of its root causes, it is important to analyze the interplay between 
inflation and speculation in a integrated manner, and this is what we set ourselves to do in this paper.  

We take as a starting point the model first proposed in \cite{Keen1995} to described the joint 
dynamics of wages, employment, and private debt. This model was fully analyzed in 
\cite{GrasselliCostaLima2012}, the main result of which is that the model exhibits essentially two distinct  
equilibrium points: a good equilibrium characterized by a finite private debt ratio and positive wage share and 
employment rate, and a bad equilibrium characterized by an infinite debt ratio and zero wages and employment. Moreover, 
both equilibria are locally stable for a wide range of parameters, implying that the bad equilibrium must be taken seriously 
as describing the possibility of a debt-induced crisis. 

The model in \cite{Keen1995}, however, has the drawback of being expressed in real terms, so any monetary phenomenon 
such as a debt-deflation spiral could only be inferred indirectly from it. This was partially remedied in \cite{Keen2013}, where 
a price dynamics is introduced alongside a thorough discussion of the endogenous money mechanism behind the dynamics of private debt. The resulting 
9-dimensional dynamical system proposed in \cite{Keen2013}, however, is exceedingly hard to analyze beyond numerical simulations, and the 
effects of the newly introduced price dynamics are difficult to infer. 

In the present paper, we first modify the basic Keen model by adopting the price dynamics proposed in \cite{Keen2013} and nothing else. The resulting system 
is still three-dimensional, so most of the analysis in \cite{GrasselliCostaLima2012} carries over to this new setting. 
We first show how the conditions for stability for the equilibrium points that were
already present for the original Keen model need to be modified to be expressed in nominal terms and include inflation or deflation regimes.
The interpretation is mostly the same but some conditions are weakened while others are strengthened
by the addition of the price level dynamics. Overall, as the numerical examples show, it can be said that money emphasizes the stable nature of
asymptotic states of the economy, both desirable and undesirable.

Next we extend the model by adding a flow of speculative money that can be used to buy existing 
financial assets. In real terms, this extension had already been suggested in \cite{Keen1995}, but was not pursued in the 
monetary model proposed in \cite{Keen2013}. To our knowledge, we present the first analysis of an extension of the Keen model 
with both inflation and speculation. 

The model is described by a four-dimensional dynamical system, from which many interesting phenomena arise. For example, even in the good equilibrium state of economy, adding speculation can change a healthy positive inflation rate into a low deflation rate. The stability of this equilibrium point also expresses the danger that
speculation represents for wage incomes and price levels.

A third and significant analytical contribution of the paper is the description of
new equilibrium points appearing with the introduction of a price dynamics. 
These points describe an economy where the employment rate approaches zero, but nevertheless deflation is sufficiently strong to 
maintain positive equilibrium nominal wages. This stylized situation has an interesting interpretation, whereby
the real economy enters a recession simultaneous with strong deflation, so that nominal wages decrease but real wages, for 
the few workers left employed, remain stable. These equilibrium points are furthermore interesting
because sufficient conditions for their local stability are weaker than for the original equilibrium points.

The final contribution is the qualitative study of the
speculation parameters on the simulated trajectories
of the monetary Keen model with speculation.
We obtain, for an interesting range of parameters, complex limit cycles
in four dimensions, which fully illustrate the rich dynamics that can arise from the 
interaction between labour markets, debt-financed investment, and speculation in financial 
markets. 

The paper is organized as follows.
In Section \ref{sec:basic keen}, we present the Keen model of \cite{Keen1995}
in its simplest, yet stock-flow consistent form, with constant price index and no Ponzi financing. In Section \ref{sec:keen model with inflation}, 
we study the extension suggested in \cite{Keen2011} with markup-led prices.
We provide a full equilibrium and stability analysis.
In particular, we introduce the new equilibria mentioned previously,
and comment on their interpretation.

In Section \ref{sec:keen model with speculation}, we introduce the speculative dimension, repeat the equilibrium analysis,
including necessary and sufficient conditions for stability, and describe the new deflationary equilibrium regime. Section \ref{sec:numerics} is dedicated to numerical computations and simulations of previously defined systems, where we provide examples of convergence to the several possible equilibrium points. We also dedicate a paragraph to the emergence of the limit cycle phenomenon described above.  

\section{The Keen model}
\label{sec:basic keen}

We recall the basic setting of \cite{Keen1995}. Denote real output by $Y$ and assume that it is related to the stock of real capital $K$ held by firms through a constant capital-to-output ratio $\nu$ according to 
\begin{equation}
\label{eq:Y=K}
Y=\frac{K}{\nu} \;.
\end{equation}
This can be relaxed to incorporate a variable rate of capacity utilization as in \cite{GrasselliNguyenHuu2014}, but we will not pursue this generalization here. Capital is then assumed to evolve according to the dynamics
\begin{equation}
\label{eq:capital}
\dot{K} = I - \delta K \;,
\end{equation}
where $I$ denotes real investment by firms and $\delta$ is a constant depreciation rate. Firms obtain funds for investment both internally and externally. Internal funds consist of net profits after paying a wage bill and interest on existing debt, that is,
\begin{equation}
\Pi=pY-W-rB,
\end{equation}
where $p$ denotes the price level for both capital and consumer goods, $r$ is the interest rate paid by firms, and $B=L-D_f$ denotes net borrowing of firms from banks, that is, the difference between firms loans $L$ and firm deposits $D_f$. Whenever nominal investment differ from profits, firms change their net borrowing from banks, that is,
\begin{equation}
\label{debt_keen}
\dot{B}=pI-\Pi.
\end{equation}
One key assumption in \cite{Keen1995} is that investment by firms is given by $I=\kappa(\pi)Y$, where $\kappa(\cdot)$ is an increasing function of the net profit share
\begin{equation}
\label{eq:profit share}
\pi=\frac{\Pi}{pY}=1-\omega-rb,
\end{equation}
where
\begin{equation}
\label{eq:omega}
\omega=\frac{W}{pY} \quad \And \quad b=\frac{B}{pY}
\end{equation}
denote the wage and firm debt shares, respectively. 
The investment function $\kappa$ is assumed to be differentiable on $\R$, 
verifying also the following technical conditions invoked in \cite{GrasselliCostaLima2012}:
\begin{eqnarray}
\label{kappa 2}& \lim\limits_{\pi\to -\infty}\kappa(\pi) = \kappa_0 < \nu(\alpha+\beta+\delta)< \lim\limits_{\pi\to +\infty}\kappa(\pi) \;,& \\
\label{kappa 3}& \lim\limits_{\pi\to -\infty}\pi^2 \kappa'(\pi) = 0 \;. &
\end{eqnarray}
Denoting the total workforce by $N$ and the number of employed workers by $\ell$, we can define the productivity per worker $a$, the employment rate $\lambda$, and the nominal wage rate as
\begin{equation}
\label{productivity}
a =\frac{Y}{\ell}, \qquad
\lambda  = \frac{\ell}{N} = \frac{Y}{aN} \;, \qquad
\mathrm{w}=\frac{W}{\ell} \;.
\end{equation}
We then assume that productivity and workforce follow exogenous dynamics of the form
\begin{equation}
\label{eq:productivity}
\frac{\dot a}{a} = \alpha, \qquad
\frac{\dot N}{N} = \beta,
\end{equation} 
for constants $\alpha$ and $\beta$, which leads to the employment rate dynamics 
\begin{equation}
\label{employment_dynamics}
\frac{\dot{\lambda}}{\lambda} = \frac{\dot{Y}}{Y} - \alpha-\beta.
\end{equation}

The other key assumption in \cite{Keen1995} is that changes in the wage rate are related to the current level of employment by a classic Phillips curve, that is, 
the wage rate evolves according to
\begin{equation}
\label{wage_dynamics}
\frac{\dot{\mathrm{w}}}{\mathrm{w}}=\Phi(\lambda),
\end{equation}
where $\Phi(\cdot)$ is an increasing function of the employment rate. 
The function $\Phi$ is defined on $[0,1)$ and takes values in $\R$.
It is assumed to be differentiable on that interval, with a vertical asymptote at $\lambda=1$, and to satisfy 
\begin{equation}
\label{phi_0}
\Phi(0)<\alpha,
\end{equation}
so that the equilibrium point defined below in \eqref{eq:omega^1}-\eqref{eq:b^1} exists. 

The model in \cite{Keen1995} assumes that all variables are quoted in real terms, which is equivalent to assume that the price level is constant and 
normalized to one, that is, $p\equiv 1$. Under this assumption, which we will relax in the next section, it is straightforward to see that the dynamics of the wage share, employment rate, and debt share reduce to the three-dimensional system 
\begin{equation}
\left\{
\begin{array}{ll}
\label{keen_original}
 \dot{\omega} &= \omega \left[\Phi(\lambda) - \alpha \right]\\
 \dot{\lambda} &= \lambda \left[g(\pi)-\alpha - \beta\right]\\
 \dot{b} &= \kappa(\pi)-\pi -bg(\pi) 
\end{array}
\right. 
\end{equation}
where
\begin{equation}
\label{eq:growth rate}
g(\pi):=\frac{\dot Y}{Y}=\frac{\kappa(\pi)}{\nu}-\delta,
\end{equation}
is the growth rate of the economy for a given profit share $\pi=1-\omega-rb$. 

The properties of \eqref{keen_original} were extensively analyzed in \cite{GrasselliCostaLima2012}, 
where it is shown that the model exhibits essentially two economically meaningful equilibrium points. 
The first one corresponds to the equilibrium profit share
\begin{equation}
\label{eq:pi}
\overline\pi_1 = g^{-1}(\alpha+\beta)=\kappa^{-1}(\nu(\alpha+\beta+\delta)),
\end{equation}
which we substitute in system \eqref{keen_original} to obtain
\begin{eqnarray}
\label{eq:omega^1}\overline\omega_1 &=& 1 - \overline\pi_1 - r \overline b_1\\
\label{eq:lambda^1}\overline\lambda_1 &=& \Phi^{-1}(\alpha) \\
\label{eq:b^1}\overline b_1 &=& \frac{\kappa(\overline\pi_1) - \overline\pi_1}{\alpha+\beta}.
\end{eqnarray}
This point is locally stable if and only if 
\begin{equation}
\label{eq:condition stability good}
0< r\left(1+ \kappa'(\overline \pi_1)\left(\frac{\overline b_1}{\nu}-1\right)\right)<\alpha+\beta \;
\end{equation}
It corresponds to a desirable equilibrium point, with finite debt, positive wages and employment rate, 
and an economy growing at its potential pace, namely with a growth rate at equilibrium equal to
\begin{equation}
\label{eq:equilibrium good growth}
g(\overline\pi_1)=\frac{\kappa(\overline\pi_1)}{\nu}-\delta =\alpha+\beta\;,
\end{equation}
that is, the sum of population and productivity growth. 

By contrast, the other equilibrium point is an undesirable state of the economy characterized by  
$(\overline\omega_2, \overline\lambda_2, \overline b_2) = (0, 0, +\infty)$, with the rate of investment converging to
$\lim_{x\to -\infty}\kappa(x) = \kappa_0$.
This equilibrium is stable if and only if
\begin{equation}
\label{eq:condition stability bad}
g(-\infty)=\frac{\kappa_0}{\nu} - \delta < r \; .
\end{equation}
The key point in \cite{GrasselliCostaLima2012} is that both conditions \eqref{eq:condition stability good} and \eqref{eq:condition stability bad} are easily satisfied 
for a wide range of realistic parameters, meaning that the system exhibits two locally stable equilibria, and neither can be ruled out a priori. 

The model in \eqref{keen_original} can be seen as a special case of the stock-flow consistent model represented by the balance sheet, transactions, and flow of funds in Table \ref{table}. To see this, observe that although \cite{Keen1995} does not explicitly model consumption for banks and households, it must be the case that total consumption satisfies
\begin{equation}
\label{consumption_keen}
C=Y-I=(1-\kappa(\pi))Y,
\end{equation} 
since there are no inventories in the model, which means that total output is implicitly assumed to be sold either as investment or consumption. In other words, 
consumption is fully determined by the investment decisions of firms. This 
shortcoming of the model is addressed in \cite{GrasselliNguyenHuu2014}, where inventories are included in the model and consumption and investment are 
independently specified. For this paper, however, we adopt the original specification \eqref{consumption_keen}. Further assuming that $r_f=r_L=r$ and that 
$p=1$ leads to the system in \eqref{keen_original}. 
\begin{table}[htd]
{\small
\begin{center}
\begin{tabular}{|l|c|cc|c|c|}
\hline
 & Households & \multicolumn{2}{|c|}{Firms} & Banks &  Sum  \\
\hline 
 {\bf Balance Sheet} &  & &  & &  \\
Capital stock &  & \multicolumn{2}{|c|}{$+pK$} &   & $+pK$ \\
Deposits & $D_h$ & \multicolumn{2}{|c|}{$+D_f$} &  $-D$&  0  \\
Loans &  & \multicolumn{2}{|c|}{$-L$} & $+L$ &   0 \\
\hline
Sum (net worth) & $X_h$ & \multicolumn{2}{|c|}{$X_f$}  & $X_b$  & $X$ \\
\hline 
\hline
{\bf Transactions} & &  current & capital &    &\\
Consumption  & $-pC_h$ & $+pC$ & & $-pC_b$   &  0 \\
Investment  & & $+pI$ & $-pI$ &   & 0  \\
Accounting memo [GDP] & & [$pY$]  & &  & \\
Wages & $+W$ & $-W$ & &   & 0 \\
Interest on deposits  & $+r_{h}D_h$ & $+r_fD_f$ & &   $-r_{h}D_h-r_fD_f$ &   0 \\
Interest on loans  &  & $-r_LL$ &  & $+r_LL$ &  0 \\
\hline
Financial Balances & $S_h$ & $\Pi$ & $-pI$ & $S_b$  & 0 \\
\hline
\hline
 {\bf Flow of Funds} & &  &    &    &\\
Change in Capital Stock & & \multicolumn{2}{|c|}{$+pI$}& & $+pI$\\
Change in Deposits & $+\dot{D}_h$& \multicolumn{2}{|c|}{$+\dot{D}_f$}&  $-\dot{D}$&   0  \\
Change in Loans & & \multicolumn{2}{|c|}{$-\dot{L}$}   &  $+\dot{ L}$ &   0 \\
\hline
Column sum & $S_h$ &  \multicolumn{2}{|c|}{$\Pi$}  &   $S_b$ & $pI$\\
Change in net worth & $\dot X_h=S_h$ &\multicolumn{2}{|c|}{$\dot X_f=\Pi+(\dot{p}-\delta p)K$}   &$\dot X_b=S_b$ & $\dot X$\\
\hline
\end{tabular}
\end{center}
\caption{Balance sheet, transactions, and flow of funds for a three-sector economy.}
\label{table}
}
\end{table}

\section{Keen model with inflation}
\label{sec:keen model with inflation}

\subsection{Specification and equilibrium points}
\label{sec:keen model equilibrium}

We follow \cite{Desai1973} and \cite{Keen2013} and consider a wage-price dynamics of the form
\begin{align}
\label{wage-1}
\frac{\dot{\mathrm{w}}}{\mathrm{w}}&=\Phi(\lambda)+\gamma i\;, \\
\label{price-1}
i&=\frac{\dot p}{p} =-\eta_p\left[1-\xi\frac{\mathrm{w}}{ap}\right] =\eta_p(\xi\omega-1)
\end{align}
for a constants $0\leq \gamma \leq 1$, $\eta_p>0$ and $\xi\geq 1$. 
The first equation states that workers bargain for wages based on the current state of the labour market as in \eqref{wage_dynamics}, 
but also take into account the observed inflation rate $i$. 
The constant $\gamma$ measures the degree of money illusion, 
with $\gamma=1$ corresponding to the case where workers fully incorporate inflation in their bargaining, 
which is equivalent to the wage bargaining in real terms assumed in \cite{Keen1995}.
The second equation assumes that the long-run equilibrium price is given by a markup $\xi$ times unit labor cost $\mathrm{w}/a$, 
whereas observed prices converge to this through a lagged adjustment of exponential form with a relaxation time $1/\eta_p$. 

The wage-price dynamics in \eqref{wage-1}-\eqref{price-1} leads to the modified system
\begin{equation}
\label{keen_inflation}
\left\{
\begin{array}{ll}
\dot\omega &= \omega\left[\Phi(\lambda)-\alpha-(1-\gamma)i(\omega)\right] \\
\dot\lambda &= \lambda\left[g(\pi)-\alpha-\beta\right]  \\
\dot b &= \kappa(\pi)-\pi-b\left[i(\omega)+g(\pi)\right]
\end{array}\right. \,,
\end{equation}
where $\pi=1-\omega-rb$ and $i(\omega)=\eta_p(\xi\omega-1)$. Observe that the introduction of the price dynamics \eqref{price-1} does not increase the 
dimensionality of the model, because the price level $p$ does not enter in the system \eqref{keen_inflation} explicitly and can be 
found separately by solving \eqref{price-1} for a given solution of \eqref{keen_inflation}. 

This model has the same types of equilibrium points as the model in Section \ref{sec:basic keen}, plus two new ones. 
We start with the original points.
Namely, defining the equilibrium profit rate 
as in \eqref{eq:pi}, substitution into \eqref{keen_inflation} leads to 
the good equilibrium
\begin{eqnarray}
\label{eq:omega^1_in} \overline\omega_1 &=& 1 - \overline\pi_1 - r \overline b_1 \\
\label{eq:lambda^1_in} \overline\lambda_1 &=& \Phi^{-1}[\alpha + (1-\gamma)i(\overline\omega_1)] \\
\label{eq:b^1_in} \overline b_1 &=& \frac{\kappa(\overline\pi_1) - \overline\pi_1}{\alpha+\beta+i(\overline\omega_1)}.
\end{eqnarray}
We therefore see that, if $\xi\geq 1/\overline\omega_1$, then $i(\overline\omega_1)=\eta_p(\xi\overline\omega_1-1)\geq 0$, that is, the model is asymptotically inflationary.
In this case, it is easy to see from \eqref{eq:lambda^1_in} that the equilibrium employment rate is higher than the corresponding values \eqref{eq:lambda^1} in the basic model. Conversely, if $\xi<1/\overline\omega_1$, then the model is asymptotically deflationary, that is $i(\overline\omega_1)< 0$, leading to lower employment rate at equilibrium. This is reminiscent of the common interpretation of the Philips curve as a trade-off between unemployment and inflation, but derived 
here as a relation holding at equilibrium, with the 
Philips curve used in \eqref{wage-1} as a structural relationship governing the dynamics of the wage rate instead. 

Inserting the expression for $\overline b_1$ into \eqref{eq:omega^1_in} we see that $\overline\omega_1$ is a solution of the quadratic equation 
$a_0\omega^2+a_1\omega+a_2=0$ where
\begin{eqnarray*}
a_0 &=& \xi \eta_p >0 \\
a_1 &=& \alpha+\beta -\eta_p(1+\xi(1- \overline\pi_1)) \\
a_2 &=& (\eta_p-\alpha-\beta)(1-\overline\pi_1)+r(\kappa(\overline\pi_1) - \overline\pi_1)
\end{eqnarray*}
We naturally seek a non-negative solution to the above equation.
For low values of $\eta_p$, 
the price level in \eqref{price-1} adjusts slowly, 
and we retrieve a behavior similar to the basic Keen model of Section \ref{sec:basic keen}. 
For example, if
\begin{equation}
\label{eq:condition for existence good eq}
\eta_p< \alpha+\beta \; \mbox{ and }  \; \overline\pi_1 \le \min(1, \nu(\alpha+\beta+\delta)),
\end{equation}
then $a_2<0$ and there is a unique positive value $\overline\omega_1$.
If, however, $\eta_p$ is large enough, that is $\eta_p>\alpha+\beta$, then 
we need to further consider the discriminant condition 
\begin{equation}
\label{eq:discriminant}
\left((\alpha+\beta - \eta_p) + \eta_p\xi(1-\overline\pi_1)\right)^2>4 \eta_p \xi r (\kappa(\overline \pi_1) - \overline \pi_1)\; .
\end{equation}
If this condition holds, provided $\overline \pi_1<1$, then $a_1>0$ and at least one solution $\overline{\omega}_1$ is non-negative, and often more than one. 
Provided the equilibrium point $(\overline\omega_1, \overline{\lambda}_1, \overline{b}_1)$ exists,
its stability is analyzed in the Section \ref{sec:stab_3_good}.

Similarly, the bad equilibrium can be found by studying the modified system $(\omega, \lambda, q)$ with
$q = 1/b$, that is,
\begin{equation}
\left\{
\label{inflation_inverse}
\begin{array}{l}
\dot{\omega} = \omega \left[\Phi(\lambda) - \alpha -(1-\gamma)i(\omega)\right]\\
\dot{\lambda} = \lambda \left[g(1-\omega - r/q) - \alpha - \beta \right]\\
\dot{q} = q \left[g(1-\omega - r/q)+i(\omega)\right] - q^2 \left[\kappa(1-\omega - r/q) - (1-\omega - r/q)\right]
\end{array}
\right. \,,
\end{equation}
for which the point $(0,0,0)$ is a trivial equilibrium corresponding to
$(\overline\omega_2, \overline\lambda_2, \overline b_2) = (0, 0, +\infty)$. 
The stability of this equilibrium is analyzed in Section \ref{sec:stab_3_bad}. 

We now focus on a new feature of the monetary model \eqref{keen_inflation},
namely the possibility that at low employment rates,
a deflation regime compensates the decrease in nominal wages.
We start with an equilibrium with non-zero wage share, zero employment, and finite debt ratio given by 
$(\overline\omega_3, 0, \overline b_3)$ where 
\begin{equation}
\label{eq:third equilibrium point}
\overline\omega_3 = \frac{1}{\xi} +\frac{\Phi(0)-\alpha}{\xi \eta_p (1-\gamma)}\;
\end{equation}
and $\overline b_3$ solves the nonlinear equation
\begin{equation}
\label{eq:debt equilibrium 3}
b \left[ i(\overline \omega_3) + g(1-\overline \omega_3 - r b)-r \right] = \kappa(1-\overline{\omega}_3 - r b)-1+\overline \omega_3 \; .
\end{equation}
Notice that
$$
i(\overline \omega_1) = \Phi(\overline \lambda_1)-\alpha > \Phi(0) - \alpha = i(\overline \omega_3) \;,
$$
so that any equilibria characterized by the wage share $\overline \omega_3$ in \eqref{eq:third equilibrium point} is asymptoticly deflationary on account of condition \eqref{phi_0}. Because of \eqref{productivity} and \eqref{eq:Y=K}, a zero employment rate implies that both output and capital vanish at equilibrium.
Following \eqref{eq:omega}-\eqref{productivity}, 
the total wage bill $W$ is null but the real wage per capita $\mathrm{w}/p$ continues to grow asymptotically 
at the same rate as the productivity, namely $\alpha$, since
\begin{equation}
\frac{\mathrm{w}}{p}=\omega a
\end{equation} 
and $\omega\to \overline\omega_3$. This situation seems artificial and must be qualified.  
When this equilibrium is locally stable, it illustrates an economic situation where the decrease in employment rate
does not lead to a real wage loss for the diminishing working force, because of the corresponding decrease in the 
general price level. The state is still bad, but expresses the possibility that
economic crises do not necessarily translate into lower average real wages. 

Moreover, we see from \eqref{eq:omega} that 
a finite $\overline b_3$ leads to $B=0$ at equilibrium too, corresponding to a slowing down of the economy as a whole, including banking activities.
As we will see with the stability analysis, and \textit{a fortiori} in the case with
speculation, this situation is unlikely to happen. The reduction in nominal wages reinforces the profit share,
and thus the expansion of credit. The eventuality of deflation favours creditors at the expense of debtors, which 
in turn fosters an increase in the debt ratio. To study the possibility of exploding debt ratio in this case, if one uses the change of variable leading to the modified system \eqref{inflation_inverse}, then $(\overline \omega_3, 0, 0)$ is an equilibrium point of that system which corresponds to the 
equilibrium $(\overline \omega_3, 0, +\infty)$ of the original system \eqref{keen_inflation}. 

We therefore see that, provided local stability holds
for one or both points characterized by \eqref{eq:third equilibrium point}, we are dealing with examples of economic crises 
related to deflationary regimes which are likely, but not necessarily, accompanied by a debt crises.

\subsection{Local stability analysis}
\label{stability_inflation}

\subsubsection{Good equilibrium}
\label{sec:stab_3_good}
Assume the existence of $(\overline\omega_1, \overline\lambda_1, \overline b_1)$
as defined in Section \ref{sec:keen model equilibrium}, 
with $\overline\omega_1>0$ and $\overline\lambda_1\in (0,1)$.
The study of local stability in system \eqref{keen_inflation}
goes through the Jacobian matrix $J_3$ given by 
\begin{equation}
\label{eq:jacobian keen inflation inversed}
\begin{bmatrix}
\Phi(\lambda) - \alpha + (1-\gamma)\eta_p (1-2\xi \omega) 	& \omega \Phi'(\lambda) & 0 \\
& & \\
-\dfrac{\lambda \kappa'(\pi)}{\nu}& g(\pi) - \alpha - \beta  & -r \dfrac{\lambda \kappa'(\pi)}{\nu}\\
& & \\
\dfrac{(b-\nu)\kappa'(\pi)}{\nu} + 1 - \eta_p \xi b & 0 & \dfrac{r\kappa'(\pi)(b - \nu)}{\nu} +r - g(\pi) - i(\omega)
\end{bmatrix}.
\end{equation}
At the equilibrium point $(\overline \omega_1, \overline \lambda_1, \overline b_1)$, with
$g(\overline{\pi}_1) = \alpha+\beta$ as stated by \eqref{eq:equilibrium good growth},
this Jacobian becomes
\begin{equation*}
J_3(\overline\omega_1, \overline\lambda_1,\overline b_1) = 
\begin{bmatrix}
K_0	& K_1 & 0 \\
-K_2& 0 & - r K_2\\
K_3-\eta_p \xi \overline b_1  & 0 &K_4
\end{bmatrix},
\end{equation*}
with the terms
\begin{equation}
\label{K_0} K_0 = (\gamma-1)\eta_p \xi \overline\omega_1 <0, \quad 
K_1 = \overline \omega_1 \Phi'(\overline\lambda_1)>0 \And
K_2 = \overline\lambda_1 \frac{\kappa'(\overline\pi_1)}{\nu}>0
\end{equation}
having well-defined signs, and the terms
\begin{equation}
\label{eq:K_3}
K_3 = \frac{\kappa'(\overline\pi_1)(\overline b_1 - \nu)}{\nu}+1 \And K_4 = rK_3 - (\alpha+\beta+i(\overline{\omega}_1))
\end{equation}
having signs that depend on the underlying parameters. 
The characteristic polynomial of $J_3(\overline\omega_1, \overline\lambda_1,\overline b_1)$ is given by
\begin{equation}
\mathbb{K}_3[X] = \left(K_4 - X\right)\left(K_2 K_1 - X(K_0 - X)\right) - rK_1K_2 \left(K_3 - \eta_p\xi \overline{b}_1\right).
\end{equation}
Factorization provides
\begin{equation*}
-\mathbb{K}_3[X] = X^3  - X^2 (K_0 + K_4)+ X \left(K_0K_4 + K_1 K_2\right) + K_1 K_2K_5\; .
\end{equation*}
with $K_5=\alpha+\beta + i(\overline{\omega}_1) - r\eta_p \xi \overline{b}_1$, which provides the necessary conditions for stability of the equilibrium point:
$$
K_0 < -K_4, \quad  -K_0 K_4 > K_1 K_2, \quad K_5>0,  
\And (K_0 + K_4)(K_0 K_4 + K_1 K_2)> K_1 K_2 K_5.
$$
A numerical test can be implemented to study such a condition. Notice that if $K_3>\eta_p \xi \overline b_1$, then $\mathbb{K}_3$ has three negative roots if
$
\left(K_4 - X\right)\left(K_2 K_1 - X(K_0 - X)\right)
$
has three negative roots.
According to \eqref{K_0}, $K_1 K_2 >0$ and $K_0<0$, so that the Routh-Hurwitz criterion for the later
polynomial reduces to the last condition $K_4<0$.
A sufficient condition for having three negative roots to $\mathbb{K}_3$ is thus
\begin{equation}
\label{eq:condition stability good 3}
r\eta_p \xi \overline b_1 <r\left(1+\kappa'(\overline \pi_1)\left(\frac{\overline b_1}{\nu}-1\right)\right)<\alpha+\beta+i(\overline{\omega}_1) \; .
\end{equation}
This condition resembles \eqref{eq:condition stability good} with modifications.
The left-hand side inequality is a stronger condition than \eqref{eq:condition stability good}
if $\overline b_1>0$, which is expected.
On the contrary, the righ-hand side inequality of \eqref{eq:condition stability good 3}
is a weaker condition if $i (\overline\omega)>0$, which is also expected.

\subsubsection{Bad equilibrium}
\label{sec:stab_3_bad}
As stated in Section \ref{sec:keen model equilibrium},
a bad equilibrium emerges if we study a modified system with state space $(\omega, \lambda, q)$ with $q=1/b$.
Assuming \eqref{kappa 3}, the Jacobian matrix of the modified system \eqref{inflation_inverse} at the point $(0,0,0)$ is
\begin{equation}
\label{eq:jacobian keen inflation value}
J_{\eps}(0,0,0) = 
\begin{bmatrix}
\Phi(0) - \alpha + (1-\gamma)\eta_p  	& 0 & 0 \\
0 & \dfrac{\kappa_0}{\nu} - \alpha - \beta -\delta & 0\\
0& 0 &  \dfrac{\kappa_0}{\nu} - \delta -\eta_p-r
\end{bmatrix}.
\end{equation}
This matrix is similar to the one found in \cite{GrasselliCostaLima2012}.
Provided \eqref{kappa 2} holds, the bad equilibrium is locally stable if and only if
\begin{equation}
\label{eq:condition bad stability}
\Phi(0) < \alpha - (\gamma-1)\eta_p \quad \And \quad \frac{\kappa_0}{\nu} - \delta -\eta_p< r\;.
\end{equation}
Notice that, especially  for high values of $\eta_p$, the first condition above is stronger than \eqref{phi_0}, which was a necessary and sufficient condition 
for stability of the bad equilibrium in the original Keen model without inflation (see \cite{GrasselliCostaLima2012}). 
On the contrary, the second condition above is weaker than condition \eqref{eq:condition stability bad},
but bares the same interpretation, with the nominal growth rate replacing the real growth rate in the comparison with the interest rate $r$.

\subsubsection{New equilibria}
\label{sec:stab_3_new}

We now study the previously computed Jacobian matrices on the new equilibria defined at the end of Section \ref{sec:keen model equilibrium}.
Take first the point $(\overline\omega_3, 0,\overline b_3)$ with $\overline\omega_3$ defined by \eqref{eq:third equilibrium point} and
$\overline b_3$ defined by \eqref{eq:debt equilibrium 3}. We obtain from \eqref{eq:jacobian keen inflation inversed}
\begin{equation}
\label{eq:jacobian keen inflation bad}
J_3(\overline\omega_3, 0,\overline b_3) = 
\begin{bmatrix}
K'_0	& K'_1 & 0 \\
0& g(\overline \pi_3)-\alpha-\beta & 0\\
K'_3-\eta_p \xi \overline b_3  & 0 & K'_4
\end{bmatrix}
\end{equation}
where $K'_0, K'_1, K'_3$ and $K'_4$ are given by \eqref{K_0} and \eqref{K_4} where $(\overline\omega_1, \overline\lambda_1, \overline b_1)$ is replaced by $(\overline\omega_3, 0, \overline b_3)$, 
and $\alpha+\beta$ in $K_4$ is replaced by $g(\overline \pi_3)=\frac{\kappa(\overline \pi)}{\nu}-\delta$, with $\overline \pi=1-\overline\omega_3-r\overline b_3$.
The Jacobian for $(\lambda, \omega, b)$ is thus lower triangular, which readily provides 
eigenvalues for $J_3(\overline\omega_3, \overline\lambda_3,\overline b_3)$, 
and the conditions for local stability:
\begin{equation}
\label{eq:local stability of 3}
(\gamma - 1)\eta_p \xi \overline \omega_3 <0, \,\, g(\overline{\pi}_3)<\alpha+\beta \ \And \ 
 r\left(1+\kappa'(\overline \pi_3)\left(\frac{\overline b_3}{\nu}-1\right)\right)<g(\overline{\pi}_3)+i(\overline \omega_3) \; .
\end{equation}
The first condition above is always satisfied. The second condition, however, fails to hold 
whenever $\overline \pi_3>\overline \pi_1$, which must be checked numerically, since \eqref{eq:debt equilibrium 3} does not have an explicit solution. The third condition is reminiscent of \eqref{eq:condition stability good} and \eqref{eq:condition stability good 3} and must also be checked numerically.

Turning to the equilibrium $(\overline\omega_3,0,+\infty)$, that is $\overline q_3=0$ in the modified system \eqref{inflation_inverse}, 
the Jacobian matrix $J_{\eps}(\overline\omega_3, 0, 0)$ at point defined by \eqref{eq:third equilibrium point} is the same as in \eqref{eq:jacobian keen inflation value}, except for one zero term changed in 
$\overline{\omega}_3 \Phi'(0)$, and the diagonal terms. 
The conditions for local stability are thus
\begin{equation}
\label{eq:local stability of 3 infinite}
(\gamma-1)\eta_p \xi \overline \omega_3 <0 \quad \And \quad \frac{\kappa_0}{\nu} - \delta +  i(\overline \omega_3) < r\;.
\end{equation}
Again, the first condition is always satisfied, 
whereas the second one should be interpreted as the comparison
between nominal growth rate and nominal interest rate, similarly to the second stability condition in \eqref{eq:condition bad stability} for 
the bad equilibrium $(0,0,+\infty)$. 
\section{Keen model with inflation and speculation}
\label{sec:keen model with speculation}

\subsection{Assumptions and equilibria} 

Borrowing for speculative purposes was modeled in \cite{Keen1995} by modifying the debt dynamics equation \eqref{debt_keen} to
\begin{equation}
\label{debt_keen_ponzi}
\dot{B}=pI-\Pi+F,
\end{equation}
where the additional term $F$ corresponded to the flow of new credit to be used solely to purchase existing financial assets. The dynamics of $F$ itself was modeled in 
\cite{Keen1995} as
\begin{equation}
\label{flow_keen}
\dot F=\Psi(g(\pi))Y,
\end{equation}
where $\Psi(\cdot)$ is an increasing function of the observed growth rate $g(\pi)$ of the economy.

In the analysis presented in \cite{GrasselliCostaLima2012}, this was changed to
\begin{equation}
\label{flow_GCL}
\dot F=\Psi(g(\pi))F,
\end{equation}
in order to ensure positivity of $F$. It was then shown that the extended system for the variables $(\omega,\lambda,b,f)$, where $f=F/Y$, admitted 
$(\overline\omega_1, \overline\lambda_1, \overline b_1,0)$ as a good equilibrium, with $\overline\omega_1$, $\overline\lambda_1$, $\overline b_1$ defined 
as in \eqref{eq:omega^1}-\eqref{eq:b^1}, but with local stability requiring that
\begin{equation}
\Psi(\alpha+\beta)<\alpha+\beta,
\end{equation}  
in addition to the previous condition \eqref{eq:condition stability good}. Moreover, \cite{GrasselliCostaLima2012} also provide the conditions for local stability 
for the bad equilibria corresponding to $(\omega,\lambda,b,f)=(0,0,+\infty,0)$ and $(\omega,\lambda,b,f)=(0,0,+\infty,+\infty)$, and showed that these were wider than the corresponding conditions in the basic Keen model. In other words, the addition of a speculative flow of the form \eqref{flow_GCL} makes it harder to achieve stability for the good 
equilibrium and easier for the bad equilibrium. 

In this paper, we revert back to the original formulation in \cite{Keen1995}, because it allows for more flexible modelling of the flow of speculative credit, which as we will see can be either positive or negative at equilibrium. In addition, in accordance with the previous section, we continue to assume a wage-price dynamics of the form 
\eqref{wage-1}-\eqref{price-1} and modify \eqref{flow_keen} to
\begin{equation}
\label{flow_keen_update}
\dot F=\Psi(g(\pi)+i(\omega))pY,
\end{equation}
where $\Psi(\cdot)$ is now an increasing function of the growth rate of nominal output. Defining the corresponding 
state variable as $f=F/(pY)$, it then follows that the model corresponds to the four-dimensional system
\begin{equation}
\label{keen_ponzi}
\left\{
\begin{array}{ll}
\dot\omega &= \omega\left[\Phi(\lambda)-\alpha-(1-\gamma)i(\omega)\right] \\
\dot\lambda &= \lambda\left[g(\pi)-\alpha-\beta\right]  \\
\dot b &= \kappa(\pi)-\pi-b\left[g(\pi)+i(\omega)\right]+f \\
\dot f &= \Psi(g(\pi)+i(\omega))-f\left[g(\pi)+i(\omega)\right]
\end{array}\right. \,.
\end{equation}

Similarly to the model \eqref{keen_inflation}, 
new equilibria emerge along with familiar ones.
With the addition of the speculative dimension $f$, we see 
that the point $(\overline\omega_1, \overline\lambda_1, \overline b_1,\overline f_1)$
obtained by defining $\overline \pi_1$ as in \eqref{eq:pi}, so that $g(\overline \pi_1) = \alpha+\beta$, and setting
\begin{eqnarray}
\label{eq:omega_1 4-dim} \overline\omega_1 &=& 1 - \overline\pi_1 - r \overline b_1 \\
\label{eq:lambda_1 4-dim} \overline\lambda_1 &=& \Phi^{-1}[\alpha + (1-\gamma)i(\overline\omega_1)] \\
\label{eq:b_1 4-dim}\overline b_1 &=&  \frac{\kappa(\overline\pi_1) - \overline\pi_1+\overline f_1}{\alpha+\beta+i(\overline\omega_1)} \\
\label{eq:f_1 4-dim} \overline f_1  &=&\frac{\Psi(\alpha+\beta+i(\overline\omega_1))}{\alpha+\beta+i(\overline\omega_1)}
\end{eqnarray}
is a good equilibrium for \eqref{keen_ponzi}.
Finding this point requires solving \eqref{eq:omega_1 4-dim}, \eqref{eq:b_1 4-dim} and \eqref{eq:f_1 4-dim} simultaneously using the 
definition of $i(\overline \omega_1)=\eta_p(\xi\overline\omega_1-1)$. Considering the change of variable $X=\alpha+\beta+\overline i(\overline \omega_1)$,
this is equivalent to solve the following equation for $X$:
\begin{equation}
\label{eq:X equation}
X^3 + \left(\eta_p \xi(\overline{\pi}_1-1) - \alpha - \beta + \eta_p\right)X^2+r\eta_p\xi\left(\kappa(\overline{\pi}_1) - \overline \pi_1\right) X + r\eta_p \xi \Psi(X) = 0 \;.
\end{equation}
Since the polynomial part of \eqref{eq:X equation} is of order three, 
it crosses the non-decreasing term $r\eta_p \xi \Psi(X)$ at least once, implying the existence of at least one solution to \eqref{eq:X equation}.
The good equilibrium satisfies $\overline{\omega}_1>0$ if and only if the corresponding solution
to \eqref{eq:X equation} verifies $X>\alpha+\beta - \eta_p$.
As we can see, provided $\alpha+\beta+i(\overline\omega_1)>0$, a positive equilibrium speculative flow $\overline f_1$ leads to a higher equilibrium borrowing ratio $\overline b_1$ and consequently lower equilibrium wage share $\overline\omega_1$ compared to the equilibrium values for the model without speculation. The stability of this equilibrium is analyzed in 
Section \ref{sec:stab_4_finite}.

Similarly to \cite{GrasselliCostaLima2012}, the change of variables  $x=1/f$ and $v=f/b$ allows us to study the bad equilibria given by
$(\overline\omega_2,\overline\lambda_2,\overline b_2,\overline f_2)=(0,0,+\infty,\pm\infty)$ and 
$(\overline\omega_2,\overline\lambda_2,\overline b_2,\overline f_2)=(0,0,+\infty,\overline{f}_{0,\infty})$ 
where 
\begin{equation}
\label{f_zer_infty}
\overline f_{0,\infty}  =\frac{\Psi\left(g(-\infty)-\eta_p\right)}{g(-\infty)-\eta_p}.
\end{equation}
There are thus two possible crisis states for the speculative flow.
One corresponding to a finite ratio $\overline f_{0,\infty}$ which corresponds to a  
financial flow $\overline F_{0,\infty}=0$ (since $Y=0$ whenever $\lambda=0$). The other corresponds to the explosion of $f$,
but at lower speed compared to $b$. We refer to \cite{GrasselliCostaLima2012} for a full interpretation. The local stability of these two types of 
equilibria are studied in Sections \ref{sec:stab_4_verybad} and \ref{sec:stab_4_bad}, respectively. 

Next we consider the equilibrium $(\overline \omega_3,0,\overline b_3,\overline f_3)$ where
$\overline \omega_3$ is given as in \eqref{eq:third equilibrium point}, $\overline b_3$ solves 

$$
b\left(\frac{\kappa(1-\overline \omega_3 - rb)}{\nu} - i(\overline \omega_3)\right) = \kappa(1-\overline \omega_3 - rb) - (1-\overline \omega_3 - rb) + 
\overline f_3\;.
$$
and
\begin{equation}
\label{f_3_infty}
\overline f_3 = \frac{\Psi\left(\frac{\kappa(1-\overline \omega_3 - rb)}{\nu} - \delta+ i(\overline \omega_3)\right)}{\frac{\kappa(1-\overline \omega_3 - rb)}{\nu} - \delta+ i(\overline \omega_3)}.
\end{equation}
As in Section \ref{sec:keen model with inflation}, 
this equilibrium must be interpreted as a bad equilibrium despite the finite values taken
by state variables. The interpretation extends to $f$, which leads to 
a financial flow $\overline F_3=\overline f_3 p \overline Y_3=0$. The stability of this equilibrium points is analyzed in Section \ref{sec:stab_4_finite}. 

The final possibilities corresponds to the equilibrium points $(\overline \omega_3,0,+\infty,\pm\infty)$ and 
$(\overline \omega_3,0,+\infty,\overline f_{3,\infty})$, where $\overline \omega_3$ is given as in 
\eqref{eq:third equilibrium point} and 
\begin{equation}
\overline f_{3,\infty} = \frac{\Psi\left(g(-\infty)+ i(\overline \omega_3)\right)}{g(-\infty)+ i(\overline \omega_3)}.
\end{equation}
The stability of these equilibria is studied in Sections \ref{sec:stab_4_bad} and \ref{sec:stab_4_verybad}. Overall, the system exhibits one good equilibrium point and 
seven different bad equilibria, confirming Tolstoy's dictum on the multiplicity of states of unhappiness. 

\subsection{Local stability Analysis} 
\label{stability_speculation}

\subsubsection{Equilibria with finite debt}
\label{sec:stab_4_finite}

The good equilibrium $(\overline \omega_1, \overline \lambda_1, \overline b_1, \overline f_1)$,
and the bad equilibrium $(\overline \omega_3, 0, \overline b_3, \overline f_3)$
are studied via system \eqref{keen_ponzi}.
The Jacobian $J_4 (\omega, \lambda, b, f)$ for this system is given 
by
\begin{equation}
\label{eq:jacobian_keen_ponzi}
\begin{bmatrix}
-(1-\gamma)\eta_p \xi \omega 	& \omega \Phi'(\lambda) & 0  & 0\\
-\dfrac{\lambda \kappa'(\pi)}{\nu}& g - \alpha - \beta & -r \dfrac{\lambda \kappa'(\pi)}{\nu} & 0\\
b\left(\dfrac{\kappa'(\pi)}{\nu}-\eta_p \xi\right)+1-\kappa'(\pi) & 0 & r \left(b\dfrac{\kappa'(\pi)}{\nu}+1-\kappa'(\pi)\right)-(g+i) & 1\\
\left(\dfrac{\kappa'(\pi)}{\nu}-\eta_p \xi\right)\left(f - \Psi'(g+i)\right) & 0 & r\dfrac{\kappa'(\pi)}{\nu}(f - \Psi'(g+i)) & -(g+ i)
\end{bmatrix}
\end{equation}
with $g=\kappa(1-\omega-rb)/\nu - \delta$ and $i = \eta_p (\xi \omega - 1)$.
At the equilibrium point $(\overline \omega_1, \overline \lambda_1, \overline b_1, \overline f_1)$,
it becomes
\begin{equation*}
\begin{bmatrix}
K_0	& K_1 & 0  & 0\\
-K_2& 0 & -r K_2 & 0\\
K_3 - \eta_p \xi \overline b_1 & 0 & K_4 & 1\\
\left(\frac{\kappa'(\overline\pi_1)}{\nu}-\eta_p \xi\right)K_6 & 0 & r\frac{\kappa'(\overline\pi_1)}{\nu}K_6 & -(\alpha +\beta +\overline i)
\end{bmatrix}
\end{equation*}
with $K_0, K_1, K_2$ and $K_3$ already defined in \eqref{K_0} and \eqref{eq:K_3}, 
and
\begin{equation}
\label{K_4}
K_6 = \overline f_1-\Psi'(\alpha+\beta+\overline i) \;.
\end{equation}
The characteristic polynomial is given non-trivially by
\begin{eqnarray*}
\label{eq:characteristic poly 4}
\mathbb{K}_4 [X] &=& \left((\alpha+\beta+\overline i) + X\right)\mathbb{K}_3 [X] + rK_5\left(\frac{\kappa'(\overline \pi_1)}{\nu}(X-K_0)X + \eta_p \xi K_1 K_2\right)\\
\nonumber &=& X^4 + a_3 X^3 + a_2 X^2 + a_1 X + a_0 \;,
\end{eqnarray*}
where
\begin{align*}
a_3 &= -K_0 - r K_3\\
a_2 &= K_0 K_4 + K_1 K_2 - (\alpha+\beta + \overline i)(K_0 + K_4) - r K_6 \frac{\kappa'(\overline \pi_1)}{\nu} \\
a_1 &= (\alpha+\beta + \overline i)K_0 K_4 + r \eta_p \xi \overline b_1 K_1 K_2 - r K_6 \frac{\kappa'(\overline \pi_1)}{\nu} K_0\\ 
a_0 &= -K_1K_2 \left((\alpha+\beta + \overline i)K_5 + rK_6 \eta_p \xi\right)
\end{align*}
with $K_i$ for $i=0$ to $4$ by \eqref{K_0},  
$K_5 = (\alpha+\beta + \overline i) - r\eta_p \xi \overline b_1$ and finally \eqref{eq:K_3}, $K_6$ by \eqref{K_4}.
The Routh-Hurwitz criterion in this case is given by
\begin{equation*}
a_i>0 \ \text{for}\ 0\le i \le 3\;, \  a_3 a_2> a_1\;, \And a_3 a_2 a_1 > a_1^2 + a_3^2 a_0 \; .
\end{equation*}
This is expected to be solved numerically only.

For the new equilibrium point $(\overline \omega_3, 0, \overline b_3, \overline f_3)$,
\eqref{eq:jacobian_keen_ponzi} becomes, after permuting the order of $\omega$ and $\lambda$ and defining $\overline \pi_3=1-\overline \omega_3-r\overline b_3$:
\begin{equation*}
\begin{bmatrix}
g(\overline\pi_3)-\alpha-\beta& 0 & 0 & 0\\
\overline\omega_3 \Phi'(0)& -(1-\gamma)\eta_p \xi \overline\omega_3  & 0  & 0\\
 0 & \overline b_3\left(\frac{\kappa'(\overline\pi_1)}{\nu}+\eta_p \xi\right)+1-\kappa'(\overline\pi_3) & r \left(\overline b_3\frac{\kappa'(\overline\pi_3)}{\nu}+1-\kappa'(\overline\pi_3)\right)-(g+\overline i_3) & 1\\
0 & \left(\frac{\kappa'(\pi)}{\nu}+\eta_p \xi\right)(\overline f_3 + \Psi'(g+\overline i_3)) & r\frac{\kappa'(\pi)}{\nu}(\overline f_3 + \Psi'(g+\overline i_3)) & -g -\overline i_3
\end{bmatrix}
\end{equation*}
The first two eigenvalues are given by 
$g(\overline \pi_3)-\alpha - \beta$
and
$(\gamma-1)\eta_p\xi \overline{\omega}_3<0$,
whereas a characteristic equation for the lower right $2\times 2$ square matrix is given by

$$
\left(\frac{r\kappa'(\overline \pi_3)}{\nu}(b-\nu)+r-g(\overline \pi_3)-i(\overline{\omega}_3) - X\right)
\left(X+g(\overline \pi_3) +i(\overline \omega_3)\right) + \frac{r\kappa'(\overline \pi_3)}{\nu}(\overline f_3-\Psi'(g(\overline \pi_3)+i(\overline{\omega}_3)  ) )
$$
The Routh-Hurwitz condition for a second order polynomial is the positivity of
all coefficients of the equation. The equilibrium point is thus locally stable if and only if
\begin{equation*}
r\left(1+\kappa'(\overline \pi_3)\left(\frac{\overline b_3}{\nu}-1\right)\right)>2(g(\overline \pi_3) + i(\overline \omega_3)) \;, \ 
g(\overline \pi_3)<\alpha + \beta\;, \ \And \ 
\Psi'(g(\overline \pi_3)+i(\overline \omega_3))>\overline f_3
\; .
\end{equation*}
The first condition recalls again \eqref{eq:condition stability good}, \eqref{eq:condition stability good 3} and 
\eqref{eq:local stability of 3}, but with a multiplier making the condition stronger here.
The second condition is also redundant and,
if $\overline \pi_3>\overline \pi_1$, it fails at this point
as for the model without speculation \eqref{keen_inflation}.
The last condition is an extra condition rendering stability even more difficult to reach.

\subsubsection{Equilibria with infinite debt and finite speculation}
\label{sec:stab_4_bad}

We make a change of variable to study the equilibria $(0, 0, +\infty, \overline f_{0,\infty})$ and $(\overline \omega_3, 0, +\infty, \overline f_{3,\infty})$
where $\overline\omega_3$, $\overline f_{0,\infty}$ and $\overline f_{3,\infty}$ are defined respectively in \eqref{eq:third equilibrium point}, \eqref{f_zer_infty}, 
and \eqref{f_3_infty}. The modification $q = 1/b$ provides the new system
\begin{equation*}
\left\{
\begin{array}{ll}
\dot\omega &= \omega\left[\Phi(\lambda)-\alpha-(1-\gamma)i(\omega)\right] \\
\dot\lambda &= \lambda\left[g(\pi)-\alpha-\beta\right]  \\
\dot q &= q\left[g(\pi)+i(\omega)\right]-q^2\left[\kappa(\pi)-\pi+f\right] \\
\dot f &= \Psi(g(\pi)+i(\omega))-f\left[g(\pi)+i(\omega)\right]
\end{array}\right. \,,
\end{equation*}
with now $\pi = 1-\omega - r/q$.
For both equilibrium points, $q=0$ and
we have that
$g = \kappa_0/\nu-\delta$ and 
$\kappa'(\overline \pi) = 0$.
Moreover, according to \eqref{kappa 3},

$$\lim_{q\to 0}\frac{\kappa'(\overline \pi)}{q^2} = \lim_{q\to 0}\frac{\kappa'(\overline \pi)}{q}=0\; .$$
The Jacobian matrix of this system, with these particular values in mind, is given by
\begin{equation*}
\begin{bmatrix}
\Phi(0)-\alpha-(1-\gamma)\eta_p (2\xi \omega-1) &\omega \Phi'(\lambda) & 0  & 0\\
-\dfrac{\lambda\kappa'(\pi)}{\nu}& \dfrac{\kappa_0}{\nu}-\delta-\alpha - \beta & 0 & 0\\
0 & 0& \dfrac{\kappa_0}{\nu}-\delta+i -r & 0\\
\left(\dfrac{\kappa'(\pi)}{\nu}-\eta_p \xi\right)(f - \Psi'(g+i)) & 0 & 0 & -\dfrac{\kappa_0}{\nu}+\delta-i
\end{bmatrix}
\end{equation*}
For the first equilibrium point, the term $\omega \Phi'(\lambda)$ disappears, 
whereas for the second equilibrium point, $-\lambda\kappa'(\pi)/\nu = 0$.
In both cases, the matrix is, up to a permutation, lower triangular, providing directly the eigenvalues and
the necessary and sufficient conditions for local stability of equilibria,
which are almost the same for both points.
For the first point, first term in the Jacobian provides the first condition in \eqref{eq:condition bad stability}, whereas the condition for the second point is 
$(\gamma-1)\eta_p\xi \overline \omega_3<0$ and is always satisfied.
Assuming that \eqref{kappa 2} holds, the other conditions for local stability reduce to
\begin{equation}
\label{eq:condition bad 4 finite speculation}
0<\frac{\kappa_0}{\nu}-\delta+ i(\overline \omega) <r \;.
\end{equation}
The second point (with positive wage share) is thus locally stable under stronger conditions than for the first one,
since $i(0)<i(\overline \omega_3)$.

\subsubsection{Bad equilibria with infinite debt and infinite speculation}
\label{sec:stab_4_verybad}
The second modification of the system is $v = f/b = q/x$ with $x=1/f$, providing
\begin{equation*}
\left\{
\begin{array}{ll}
\dot\omega &= \omega\left[\Phi(\lambda)-\alpha-(1-\gamma)i(\omega)\right] \\
\dot\lambda &= \lambda\left[g(\pi)-\alpha-\beta\right]  \\
\dot v &= vx \Psi(g(\pi) + i(\omega)) - v^2[x(\kappa(\pi) - \pi)+1] \\
\dot x &= x[g(\pi) + i(\omega) ] -x^2 \Psi(g(\pi)+i(\omega))
\end{array}\right. \,,
\end{equation*}
which now exhibits two equilibria: $(0,0,0,0)$ and $(\overline \omega_3, 0,0,0)$.
The first one, similarly to \cite{GrasselliCostaLima2012}, 
corresponds to a bad equilibrium with explosive debt and explosive rate of investment into pure finance.
Notice that debt rate $b$ grows much faster than the financial investment rate $f$, so that the explosive debt
corresponds to a Ponzi scheme into both real and financial sectors of the economy.
The second point bares the same interpretation, with positive wages
sustained by deflation.

In both cases, $\lambda=v=x=0$, $g(\overline \pi) = \kappa_0/\nu - \delta$ and
from \eqref{kappa 3},
\begin{equation*}
\lim_{|v|+|x|\to 0}\kappa'\left(1-\omega - \frac{r}{vx}\right) \left(\left|\frac{1}{v}\right| + \left|\frac{1}{v^2}\right| + \left|\frac{1}{x}\right| + \left|\frac{1}{vx}\right| + \left|\frac{1}{xv^2}\right|+\left|\frac{1}{vx^2}\right|\right) = 0 \; .
\end{equation*}
The Jacobian matrix at both points takes the form
\begin{equation*}
\begin{bmatrix}
\Phi(0)-\alpha-(1-\gamma)\eta_p (2\xi \omega-1) 	& \omega \Phi'(0) & 0  & 0\\
0 & \dfrac{\kappa_0}{\nu} - \delta - \alpha - \beta & 0 & 0\\
0 & 0 & -r & 0\\
0 & 0 & 0 & g + i
\end{bmatrix}
\end{equation*}
For the local stability of point $(0,0,0,0)$,
we need the first condition in \eqref{eq:condition bad stability}, whereas
similarly to above, the condition $(\gamma-1)\eta_p \xi \overline \omega_3<0$
for local stability of $(\overline \omega_3,0,0,0)$ is always satisfied.
Condition \eqref{kappa 2} provides the second condition, $r\ge 0$ the third,
and
$$
\frac{\kappa_0}{\nu}-\delta + i(\overline \omega)<0
$$
expresses the last condition.
Notice two things.
First, that $i(\overline \omega_3)>i(0)$ so that as previously, 
this last condition
holds for a wider range of parameters for the point $(0,0,0,0)$.
Second, similarly to \cite{GrasselliCostaLima2012}, the condition is
partially complementary to \eqref{eq:condition bad 4 finite speculation},
so that if the bad equilibrium with finite speculation is not locally stable,
then the bad equilibrium with infinite speculation has most chances to be.

\section{Numerical Simulations and Qualitative Analysis}
\label{sec:numerics}

\subsection{Parameters, equilibria and stability}
\label{sec:equilibrium}

\subsubsection{Basic Keen Model}
\label{sec:parameters basic keen}

We take most values from previous work, see \cite{Keen1995} and \cite{GrasselliCostaLima2012}.
We choose the fundamental economic constants to be
\begin{equation}
(\alpha, \beta, \delta, \nu, r) = (0.025, 0.02, 0.01,0.03) \;.
\end{equation}
The Phillips curve is chosen to be
\begin{equation}
\label{eq:Philips}
\Phi(\lambda) = \frac{\phi_1}{(1-\lambda)^2} - \phi_0
\end{equation}
with
$$
(\phi_0, \phi_1) = \left(\frac{0.04}{1-0.04^{2}} , \frac{0.04^{3}}{1-0.04^{2}}\right)
$$
so that $\Phi(0.96)=0$ and $\Phi(0) = -0.04$.
The investment function is given by
\begin{equation}
\label{eq:kappa}
\kappa(\pi) = \kappa_0 + \exp(\kappa_1 + \kappa_2 \pi)
\end{equation}
with 
$$
(\kappa_0, \kappa_1, \kappa_2) = (-0.0065, -5, 20)\;.
$$
The required conditions \eqref{kappa 2} and \eqref{kappa 3} are satisfied.
We recall results of \cite{GrasselliCostaLima2012}:
there exists only one reachable good equilibrium
$$
(\overline{\omega}_1, \overline \lambda_1, \overline b_1) = (0.8361, 0.9686, 0.0702)
$$
which is locally stable, because condition \eqref{eq:condition stability good} is satisfied. We notice for later use that, according to \eqref{eq:pi}, the profit share corresponding to this
equilibrium is
\begin{equation}
\label{eq:value pi}
\overline{\pi}_1= 0.1618 \;. 
\end{equation}
Moreover, the bad equilibrium 
$(\overline\omega_2,\overline \lambda_2,\overline b_2)=(0,0,+\infty)$ is locally stable too, because \eqref{eq:condition stability bad} is also satisfied.

In addition to parameters present in the basic Keen model of Section \ref{sec:basic keen},
we need to choose the parameters $\eta_p, \xi$ and $\gamma$
for the model with inflation of Section \ref{sec:keen model with inflation},
 and the function $\Psi$ for the model with speculation of Section \ref{sec:keen model with speculation}.

\subsubsection{Keen Model with inflation}
\label{sec:parameters inflation}

We recall that the parameter $\xi$ determines whether the good equilibrium is asymptotically inflationary ($\xi\overline{\omega}_1>1$) or deflationary
 ($\xi\overline{\omega}_1<1$). The price level is asymptotically constant if
$$
\xi = 1 + \frac{1-\overline \omega_1}{\overline \omega_1},
$$
where a first proxy for $1-\overline \omega_1$, 
ignoring interest payments, 
is given by $\overline \pi_1$, which is directly given
as a function of exogenously fixed parameters in \eqref{eq:value pi}. 
Replacing in the equation provides $\xi=1.193$, and we choose 
a slightly higher value to ensure an inflationary state at equilibrium, while at the same time consistent with empirical estimates (see \cite{martins1996mark}). 
The parameter $\eta_p$, on the other hand, reflects the speed of adjustment of prices to their long-term target.
We already mentioned the importance of this parameter in Section \ref{sec:keen model with inflation}.
To distinguish from the non-monetary Keen model analyzed in \cite{GrasselliCostaLima2012}, we 
take $\eta_p=4$, representing an average period of adjustment of three months. 
We thus take additional parameters as follows:
\begin{equation}
\label{eq:param inflation}
(\eta_p, \xi, \gamma) = (4, 1.2, 0.8)\;.
\end{equation}
According to this parametrization, we obtain two positive roots to the equation $a_0 \omega^2 + a_1 \omega+ a_2=0$, namely
\begin{equation}
\left\{
\begin{array}{l}
\overline{\omega}_1^+ = 0.8372\\
\overline{\omega}_1^- = 0.8250
\end{array}
\right.,
\end{equation}
corresponding to the two equilibrium points
\begin{equation}
\label{good_eq_inflation_numerical}
(\overline \omega_1^+ , \overline \lambda_1^+, \overline b_1^+)
=(0.8372,0.9695,0.0498)
\And
(\overline \omega_1^- , \overline \lambda_1^-, \overline b_1^-)
=(0.8250,0.9665,0.6602) \; .
\end{equation}
The first point leads to an inflation rate of $\overline{i}^+ = 1.838\%$, whereas the second
leads to $\overline{i}^- = -4.022\%$, corresponding respectively to higher and lower 
equilibrium employment rates compared to the base case $\overline\lambda_1=0.9686$, consistently 
with the observation following equations \eqref{eq:omega^1_in}-\eqref{eq:b^1_in}.
The local analysis formulas described in Section \ref{sec:stab_3_good}
assert, however, that in this case that only the first point is locally stable.

We also test the new bad equilibrium with finite wage share, finite debt, but zero employment rate.
Computations of \eqref{eq:third equilibrium point} provide
$\overline \omega_3 = 0.7656$.
Solving \eqref{eq:debt equilibrium 3} provides $\overline b_3=-0.9539$.
The stability analysis pursued in Section \ref{stability_inflation} shows that
this point is, as expected, locally unstable.
Indeed, the equilibrium profit share in this case is $\overline \pi_3 = 0.2535$,
which implies a very high investment share $\kappa(\overline \pi_3)$.
Accordingly, the second condition of \eqref{eq:local stability of 3} fails to hold.

The stability of the two possible bad equilibria with infinite debt is also easy to analyze.
With deflation given by $i(\overline \omega_3)<0$ we have that
\begin{equation}
\label{eq: failed condition}
\frac{\kappa_0}{\nu}- \delta  + i(\overline \omega_3)< r \;,
\end{equation}
so that the second condition in \eqref{eq:condition bad stability} holds and 
$(\overline \omega_3, 0, +\infty)$ is locally stable. However, with the 
parametrization $\eta_p = 4$ and $\gamma=0.8$, the first condition
of \eqref{eq:condition bad stability} fails holds, and consequently $(0,0,+\infty)$ is not locally stable.

\subsubsection{Keen Model with inflation and speculation}
\label{sec:parameters speculation}

To complete the model, we define the function $\Psi$ as in \cite{GrasselliCostaLima2012}:
\begin{equation}
\label{eq:Psi function}
\Psi(g) = \psi_0 (e^{\psi_2 (g-\psi_1)}-1)\;,
\end{equation}
with

$$
(\psi_0, \psi_1, \psi_2)=(0.25, 0.02, 1.2)\;.
$$
The equilibrium is found by solving \eqref{eq:X equation} numerically.
The algorithm converges to a unique solution 
and provides
\begin{equation}
\label{eq:numerical good equilibrium point}
(\overline{\omega}_1, \overline{\lambda}_1, \overline{b}_1, \overline{f}_1)
 = (0.8303,0.9679,0.2635,0.0049)\;,
\end{equation}
with the corresponding inflation rate equal to $\overline{i}=-1.18\%$. According to the criterion presented in Section \ref{stability_speculation}, the good equilibrium
is locally stable. 
Moreover, since $\overline f_1$ is positive, a regular flow of new credit goes into speculative finance at equilibrium and, as mentioned in Section \ref{sec:keen model with inflation},
the debt share is higher in \eqref{eq:numerical good equilibrium point} than in the first equilibrium in \eqref{good_eq_inflation_numerical} and the wage share is lower, as expected. 
This is an interesting result mentioned in the introduction, namely introducing speculation in the system turns an inflationary equilibrium into a deflationary one.  

The study of alternative equilibria with $\overline\omega_3$ given by \eqref{eq:third equilibrium point}
provides the following equilibrium point with finite debt and finite speculation flow:
\begin{equation}
\label{eq:numerical bad equilibrium point}
(\overline{\omega}_3, 0, \overline{b}_3, \overline{f}_3)
 = (0.7656, 0, -0.6820, 0.1006)\;,
\end{equation}
with asymptotic inflation rate $i(\overline \omega_3) = -32.50\%$. For the same reasons as without speculation,
the local stability condition does not hold for the point with finite
debt share, thus remaining a meaningless case.

Concerning the bad equilibria with infinite debt ratio, we numerically obtain with the inflation rate found above that
$$
\frac{\kappa_0}{\nu}-\delta + i(\overline \omega_3)<0 \;,
$$
which implies the local stability of the bad equilibria with infinite debt ratio and infinite speculation,
and the instability of the ones with infinite debt ratio and finite speculation.
Additionally, the failure of first condition in \eqref{eq:condition bad stability} remains here,
thus implying that the equilibrium point with zero wage share is also unstable. Consequently, 
the only bad equilibria that are locally stable in this example are 
given by $(\overline \omega_3,0,+\infty,\pm\infty)$, that is, corresponding to 
a positive wage share $\overline \omega_3$, zero employment,  
infinite debt ratio, and infinite speculation flow ratio. 

\subsection{Dynamics and behavior}
\label{sec:dynamics and behavior}

\subsubsection{Keen model with inflation}

As emphasized  in \cite{GrasselliCostaLima2012} and recalled in Section \ref{sec:basic keen},
the basic Keen model possesses two explicit locally stable equilibrium points, for a wide range of parameters.
Numerical simulations in \cite{GrasselliCostaLima2012} also show 
that no apparent strange attractor can be exhibited: the phase space
is numerically divided into two complementary regions being the basins of attraction of the
good and the bad equilibrium respectively.

The Keen model with inflation \eqref{keen_inflation}
adds three additional parameters that were fixed in \eqref{eq:param inflation}
in the previous subsection.
Changes in these parameters have the following effects.

First it appears that $\eta_p$ and $(1-\gamma)$ have,
as expected, a dampening effect on oscillations of the system.
Recall that $\eta_p$ is the speed of adjustment of prices
for a given wage share, whereas $\gamma$ is the proportion
of inflation taken into consideration of wage negotiation.
They also alter the value of the good equilibrium point. It is expected that the local basin of attraction is also affected,
but this is not pursued here. Figure \ref{fig:effect_of_price} below illustrates the dampening effect of $\eta_p$ and $(1-\gamma)$ on the 
oscillatory behavior of the employment rate $\lambda$. 
\begin{figure}[!ht]
\centering
\caption{Convergent trajectories for variable $\lambda(t)$ in system \eqref{keen_inflation} with initial
values $(\omega_0, \lambda_0, b_0)=(0.9, 0.9, 0.3)$, and 
parameters defined in Section \ref{sec:numerics}, except $(\eta_p, \gamma)$.}
\label{fig:effect_of_price}
\includegraphics[width=0.9\textwidth, height=8cm]{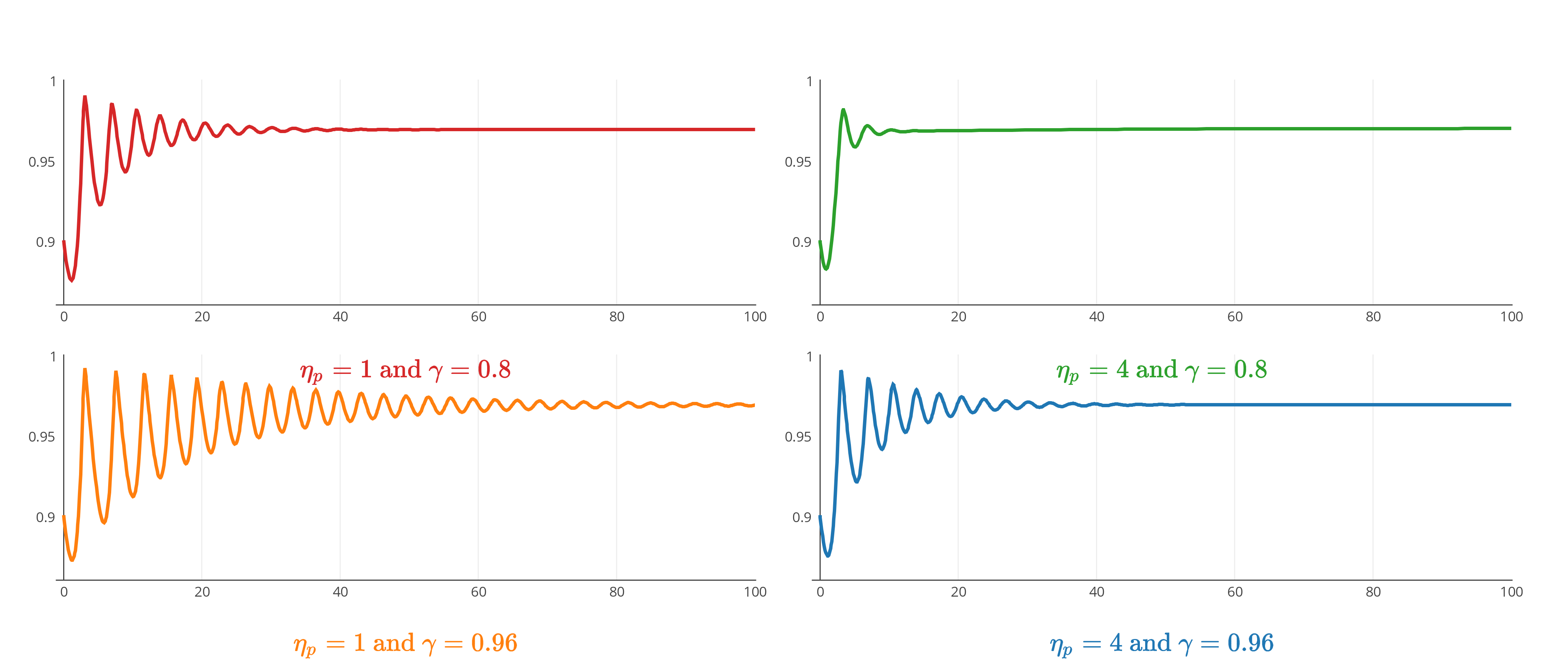}
\end{figure}

\begin{figure}[!ht]
\centering
\caption{Trajectories for variable $\omega(t)$ in system \eqref{keen_inflation} with initial
values $(\omega_0, \lambda_0, b_0)=(0.9, 0.9, 0.3)$, and 
parameters defined in Section \ref{sec:numerics}, except $\xi$.}
\label{fig:effect_xi_on_existence}
\includegraphics[width=0.9\textwidth, height=8cm]{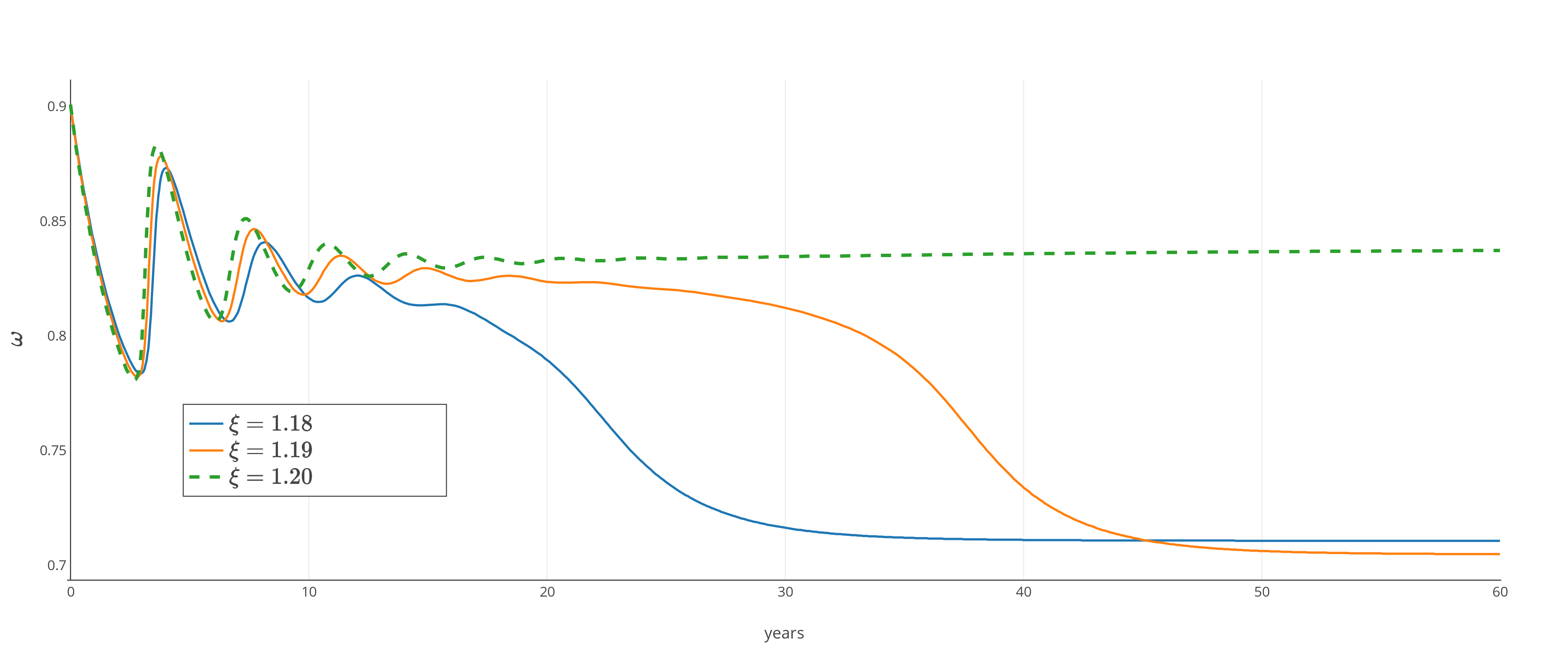}
\end{figure}

The parameter $\xi$ is even more influential.
Its value crucially impacts condition \eqref{eq:discriminant},
so that a slight decrease in the markup can lead to 
the absence of a good equilibrium point.
This is illustrated by Figure \ref{fig:effect_xi_on_existence},
where a convergence toward the bad equilibrium $(\overline \omega_3, 0, +\infty)$
appears after dampened oscillations that first seem to converge to
a good equilibrium.

The three-dimensional monetary Keen model \eqref{keen_inflation} is then
a privileged framework for studying the isolated effect
of price parameters in comparison with the basic Keen model \eqref{keen_original}.
While it exhibits new undesirable economic situations,
it also shows accelerated convergence to equilibrium points, 
thus rendering the economic situation more stable, 
whether or not the result is considered good or bad.

\subsubsection{Keen model with inflation and speculation}

\begin{figure}[!ht]
\centering
\caption{A convergent trajectory for system \eqref{keen_ponzi}, 
$(\omega_0, \lambda_0, b_0, f_0)=(0.85, 0.85, 0.5, 0.1)$, and 
parameters defined in Section \ref{sec:numerics}.}
\label{fig:convergent}
\includegraphics[width=0.9\textwidth, height=8cm]{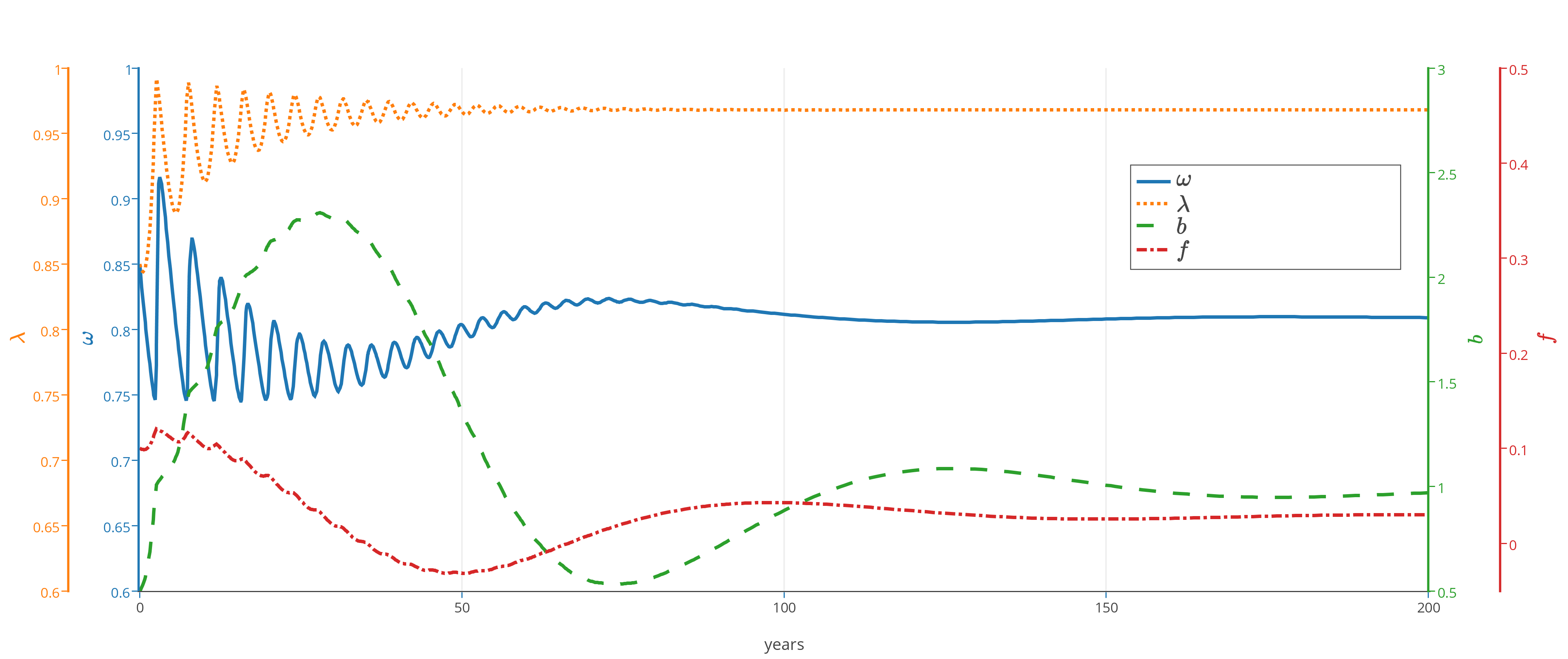}
\end{figure}

Similarly to the three-dimensional case,
the higher the value of price parameters $\eta_p$ and $(1-\gamma)$,
the faster the dampening of oscillations,
when the economy converges to the good equilibrium 
$(\overline \omega_1, \overline \lambda_1, \overline b_1, \overline f_1)$.
When we take low values for $\eta_p$ in particular,
the case is very similar to the Keen model with speculation of \cite{GrasselliCostaLima2012},
and Figure \ref{fig:convergent} pictures such a trajectory for system \eqref{keen_ponzi}
with parameter $\eta_p = 0.4$. One can then see two scales of oscillations due to different factors.
The short-period oscillations are already present in system \eqref{keen_original}:
they are given by the wage-employment variations, interpreted as business cycles in
the Goodwin model, and dampened on the long run by the use of credit to
compensate capital availability for work.
The dampening speed is also highly influenced by the relaxation parameter $\eta_p$,
which adjusts the price of capital and nominal growth to sustain wages requirements.
The long-run variations are due to the adjustment of the financial flow $f$.
The higher the parameter $\psi_2$, the shorter and the wider those fluctuations.
This parameter represents the sensitivity of the flow $F$ to
nominal growth output.

Let us now come back to the initial parametrization \eqref{eq:param inflation},
with $\eta_p=4$.
The interest here is to analyze the effect of the speculation function $\Psi$,
which is not explicit in the local stability analysis pursued in Section \ref{sec:stab_4_finite}.
The form of \eqref{eq:Psi function} implies that
$\Psi>0$ if and only if $g(\pi)+i(\omega)>\psi_1$, 
so that this last term represent the threshold of speculation direction (in or out
of financial markets).

\begin{figure}[!ht]
    \centering
    \caption{A trajectory starting near the good equilibrium point and converging to a limit cycle.
    Left: projection in the phase subspace $(\lambda, b, f)$. Right: projection in the phase subspace $(\omega, b, f)$.}
    \label{fig:limit cycle}
    \begin{subfigure} 
        \centering
        \includegraphics[width=0.45\textwidth, height=6.5cm]{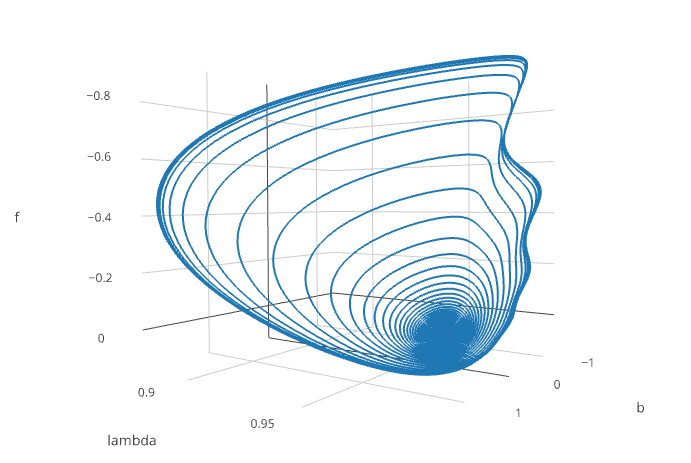}
    \end{subfigure}
    \begin{subfigure} 
        \centering
        \includegraphics[width=0.45\textwidth, height=6.5cm]{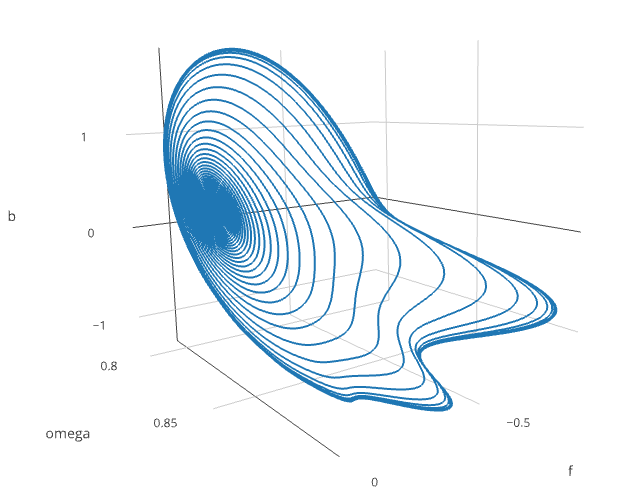}
    \end{subfigure}
\end{figure}

Depending on the value of the parameter $\psi_1$,
the system exhibits either local stability of the good equilibrium
or an absorbing limit cycle, provided that we start from
a "good" initial state $(\omega_0, \lambda_0, b_0, f_0)$.
Even when the good equilibrium point \eqref{eq:numerical good equilibrium point}
is still theoretically locally stable, its basin of attraction is very small.
Starting numerically close to that good equilibrium, the trajectory actually
converges on a very long term (+1000 years) to an elaborate limit cycle.
Figure \ref{fig:limit cycle} illustrates such a claim.

\begin{figure}[!ht]
\centering
\caption{Description of the limit cycle over time for state variables $(\omega, \lambda, b, f)$, 
and auxiliary variables $(i, g)$.}
\label{fig:limitcycle_expl}
\includegraphics[width=0.9\textwidth, height=8cm]{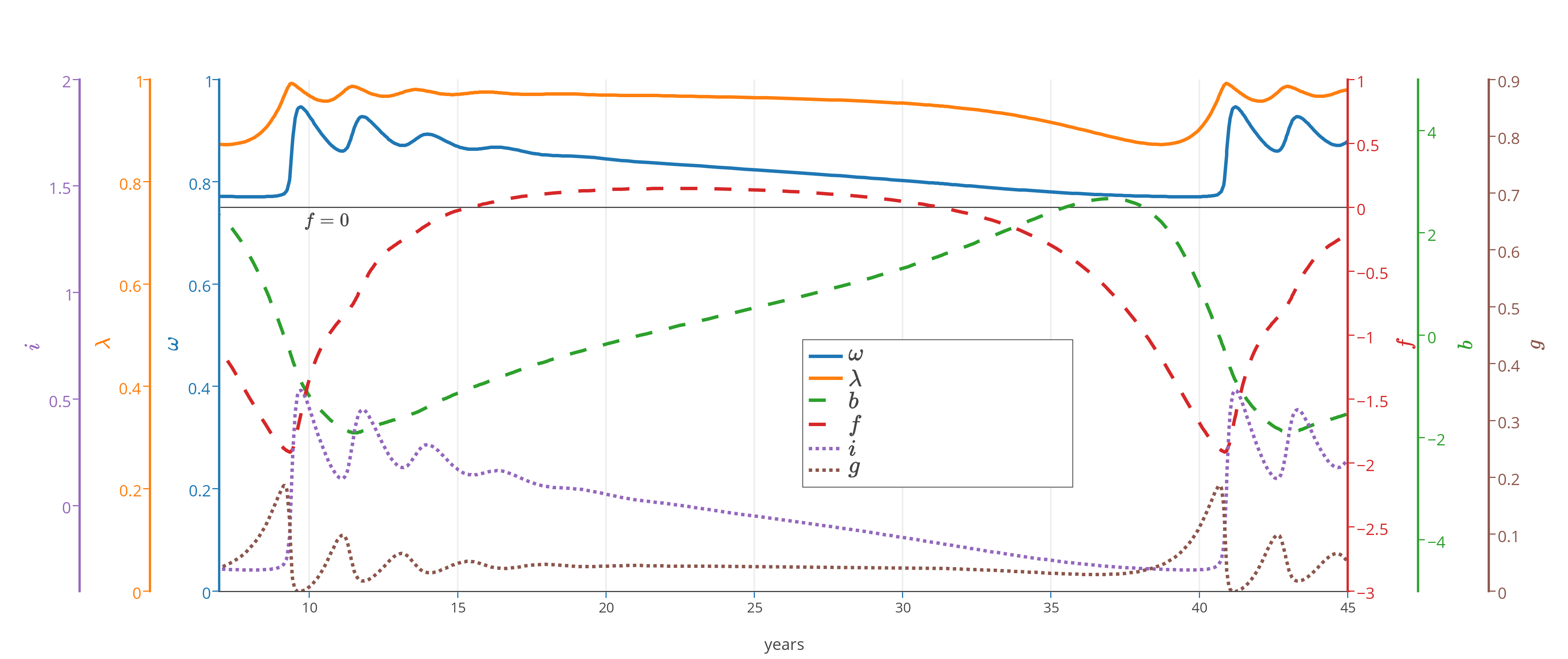}
\end{figure}

The limit cycle can be described as follows, with assistance of Figure \ref{fig:limitcycle_expl}.
Starting the economy in a state of low debt (around years 9 and 10),
it initially evolves with small cycles of oscillation between
periods of high employment rate with high wage share, and periods with
high profit share and low inflation rate. During this period, fluctuations in inflation and real growth 
cancel each other and the nominal growth has a steady, albeit slowly path. 
Speculation reacts accordingly and reduces its flow towards financial markets, directing it 
towards debt reduction instead. This in turn implies high profit rates, and pushes up the level of growth, as well as debt creation for investment purposes.
The stability of growth above a certain threshold turns the financial flow back toward financial markets, 
increasing the debt level and debt service charge and stabilizing employment and speculation flow
before year 20. The wage share steadily decreases together with inflation.
Growth remains strong until year 30, sustained by credit creation. Nominal growth 
is however impacted by deflation, which reduces the flow of speculation into financial markets again.
When the debt burden is too high, real investment decreases,
reducing the output growth. At this point the speculative flow decreases sharply, 
and the debt burden decreases and allows for high profits again, high employment rate
and in consequence, higher wages, around year 40.

If we reduce the value of $\psi_1$ in the above parameters configuration,
then the speculation flow remains positive for smaller values of the nominal growth rate.
There is no reversal of flow from speculative purposes back into debt repayment, and the debt service charges remain high.
What happens is that a trajectory starting from the previous limit cycle then converges to
the bad equilibrium $(\overline \omega_3, 0, +\infty, +\infty)$.

\begin{figure}[!ht]
\centering
\caption{A trajectory converging to the bad equilibrium $(\overline \omega_3, 0, +\infty, +\infty)$, 
starting from a point on the limit cycle of Figure \ref{fig:limitcycle_expl}.}
\label{fig:divergent}
\includegraphics[width=0.9\textwidth, height=7cm]{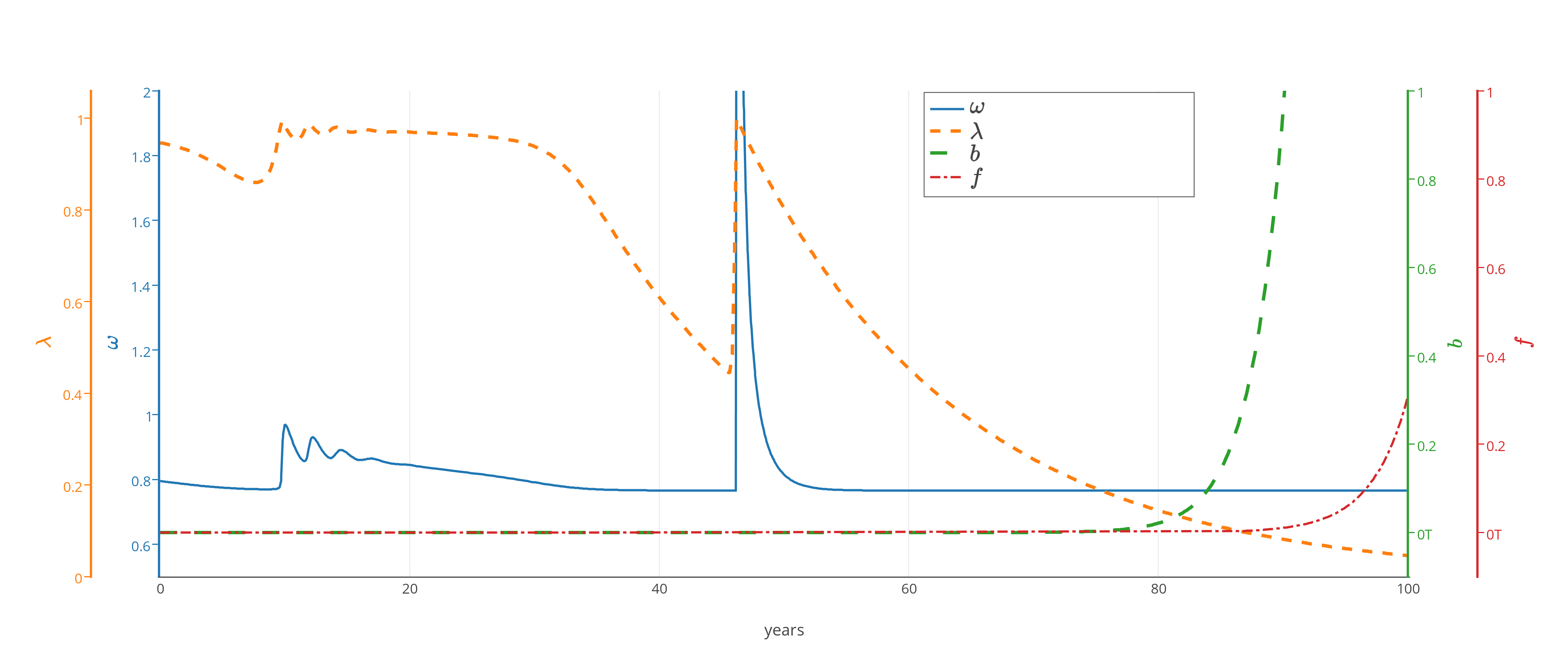}
\end{figure}

\bibliography{finance}
\bibliographystyle{plain}

\end{document}